\documentclass[journal,comsoc]{IEEEtran}
\usepackage[outdir=./]{epstopdf}
\usepackage{float}
\usepackage{caption}
\usepackage{subcaption}
\usepackage[ruled,vlined,algo2e]{algorithm2e}

\SetKwInput{KwInput}{Input}                % Set the Input
\SetKwInput{KwOutput}{Output} 
\usepackage{array}
\newcolumntype{P}[1]{>{\centering\arraybackslash}p{#1}}
\usepackage{textcomp}
\usepackage{sansmathaccent}
\usepackage{babel,blindtext}
\usepackage{graphicx} % remove [demo] in your file
\usepackage{lipsum}
\usepackage[table]{xcolor}
\usepackage{multirow}
\usepackage{graphicx}
\usepackage{grffile}
\usepackage{float}
\usepackage{upgreek}
\usepackage{filecontents,lipsum}
\usepackage[noadjust]{cite}
\usepackage{wrapfig}
\usepackage{longtable}
\usepackage{graphicx}
\usepackage{multicol}
\usepackage{algorithm}
\usepackage{algorithmic}
\usepackage{booktabs}
\usepackage[outdir=./]{epstopdf}
\usepackage{url, longtable, enumerate, setspace,amsmath, listings, color}
\DeclareCaptionLabelSeparator{periodspace}{.\quad}
\usepackage{graphicx}
\usepackage{caption}
\usepackage{amsmath}
\usepackage{amssymb}
\usepackage{amsthm}
\usepackage{optidef}

\usepackage{tikz}

\usepackage{amssymb}% http://ctan.org/pkg/amssymb
\usepackage{pifont}% http://ctan.org/pkg/pifont
\usepackage{array,booktabs}
\newcolumntype{M}[1]{>{\centering\arraybackslash}m{#1}}
\usepackage{hyperref}
\usepackage{cleveref}
\ifCLASSINFOpdf
\usepackage{amsmath}
\usepackage[cmintegrals]{newtxmath}

\title{Tri-Level Model for Hybrid Renewable Energy Systems}

\author{Eghbal Hosseini$^{1}$
	
\thanks{$^{1}$Institute of Informatics and Computing in Energy (IICE), UNITEN, 43000 Kajang, Selangor, Malaysia}
    	
}

\begin{document}
	\maketitle
\begin{abstract}
In practical scenarios, addressing real-world challenges often entails the incorporation of diverse renewable energy sources, such as solar, energy storage systems, and greenhouse gas emissions. The core purpose of these interconnected systems is to optimize a multitude of factors and objectives concurrently. Hence, it is imperative to formulate models that comprehensively cover all these objectives. This paper introduces tri-level mathematical models for Hybrid Renewable Energy Systems (HRESs), offering a framework to concurrently tackle diverse objectives and decision-making levels within the realm of renewable energy integration. The proposed model seeks to maximize the efficiency of solar PV, enhance the performance of energy storage systems, and minimize greenhouse gas emissions.
\end{abstract}


\section{Introduction}
In the realm of optimization, meta-heuristic algorithms have garnered significant attention as versatile tools applicable across a spectrum of domains demanding efficient optimizers \cite{1, 2, 3, 4, 5, 6, 7, 8, 9, 10}. Complementarily, heuristic algorithms have also captured substantial interest as optimization tools in diverse applications that necessitate efficient optimizers \cite{11, 12, 13, 14, 15, 16, 17, 18, 19, 20, 21, 22, 22-1, 22-2, 22-3, 22-4, 22-5}. The collective exploration of these algorithmic approaches reflects the expansive and evolving landscape of optimization techniques in addressing complex challenges across various disciplines\cite{23, 24, 25, 26, 27, 28, 29, 30, 31, 32, 33, 34, 35}.\par
In recent times, there has been a remarkable upswing in the advancement of various formulations specifically designed for modeling HRES. These cutting-edge formulations and mathematical models have been extensively documented in a wide range of references \cite{36, 37, 38, 39, 40, 41, 42, 43, 44}. Furthermore, there has been a proliferation of varied methodologies aimed at optimizing HRES \cite{45, 46, 47, 48, 49, 50, 51}.\par
The tri-level optimization problem is characterized by the presence of three distinct levels of objectives, each with its own set of constraints. Researchers have made several endeavors to formulate tri-level optimization problems across various applications, as evidenced by a range of studies spanning from \cite{52, 53, 54, 55, 56, 57, 58, 59, 60}. This multifaceted problem structure necessitates a nuanced approach, and the literature reflects ongoing efforts to address and apply tri-level optimization across diverse domains.\par
Moreover, the exploration of tri-level models extends into the domain of energy applications, as evidenced by a series of proposed models documented in references \cite{61, 62, 63, 64, 65, 66, 67, 68, 69, 70, 71, 72, 73, 74, 75, 76, 77, 78}.\par
A substantial body of research is dedicated to modeling and solving challenges associated with solar power, energy storage, and greenhouse gas emissions. A comprehensive review of the literature reveals a wealth of knowledge, with numerous studies contributing to the understanding and advancement of these critical aspects. A plethora of references, spanning from \cite{79, 80, 81, 82, 83, 84, 85, 86, 87, 88, 89}, underscores the extensive exploration and analysis conducted by researchers in this domain.\par
Researchers have delved into intricate models that encapsulate the dynamics of solar power systems, seeking to enhance efficiency, output, and integration within broader energy frameworks. These studies often incorporate innovative technologies and methodologies, reflecting the dynamic nature of solar energy research in the quest for sustainable solutions.\par
Energy storage, a linchpin in the transition to renewable energy, has also been a focal point of investigation. The literature showcases endeavors to optimize the performance, reliability, and economic feasibility of energy storage systems, with a particular emphasis on addressing the intermittency challenges inherent in renewable energy sources.\par
In tandem, greenhouse gas emissions reduction strategies have garnered significant attention. Researchers have proposed diverse models aimed at quantifying, mitigating, and monitoring emissions, with an overarching goal of fostering environmentally responsible energy practices. These studies contribute valuable insights to the ongoing discourse surrounding climate change mitigation and sustainable energy development.\par
As the references from \cite{90, 91, 92, 93, 94, 95, 96, 97, 98, 99, 100, 101, 102, 103} attest, this collective body of research serves as a robust foundation for understanding, modeling, and solving the intricate challenges posed by solar power, energy storage, and greenhouse gas emissions. The diversity of perspectives and methodologies within this range of references underscores the interdisciplinary nature of sustainable energy research, showcasing the collaboration between experts from various fields to address the multifaceted dimensions of our global energy transition.

As the global demand for sustainable energy solutions continues to rise, the integration of diverse renewable energy sources has become a paramount consideration in addressing real-world challenges. In contemporary energy landscapes, the urgent need to transition towards sustainable practices has led to a surge in interest and investment in renewable energy technologies. The integration of solar power, energy storage systems, and the reduction of greenhouse gas emissions are pivotal aspects of this transition. Real-world challenges demand a comprehensive approach that optimizes various factors simultaneously, requiring sophisticated models to address the complexity of these systems. This paper focuses on the development and application of mathematical models to optimize HRESs, which combine various sources such as solar, energy storage systems, and measures to mitigate greenhouse gas emissions.
\section{Tri-level for HRESs}
Hybrid renewable energy systems stand as beacons of innovation in the realm of sustainable energy technology, harnessing the collective potential of diverse renewable sources to accomplish a myriad of objectives. These pioneering systems not only represent a technological breakthrough but also play a pivotal role in our collective global commitment to mitigating the adverse effects of climate change by curbing greenhouse gas emissions.\par
At the heart of their significance is the capability to synergize different renewable sources, such as solar, wind, and energy storage systems, into an integrated and harmonious energy framework. By seamlessly blending these sources, hybrid systems enhance the overall efficiency and reliability of energy production, ensuring a more resilient and sustainable energy infrastructure.\par
In addition to their environmental benefits, hybrid renewable energy systems contribute to energy independence and security by diversifying the energy mix. This diversification helps mitigate the vulnerabilities associated with relying solely on conventional energy sources. Furthermore, these systems foster technological advancements, encouraging ongoing research and development in the pursuit of more efficient and cost-effective solutions.\par
As we navigate the complex landscape of energy transition, hybrid renewable energy systems emerge as not just technological marvels but as catalysts for positive change. Their implementation signifies a transformative shift towards a cleaner, more sustainable energy future, aligning with our shared aspirations for a planet powered by greener and more resilient energy systems.\par
Introducing a cutting-edge and meticulously crafted optimization model, we present the tri-level framework meticulously tailored for HRESs. This innovative model is strategically designed to enhance the efficiency and performance of grid-connected HRESs. The primary focus of our model revolves around the seamless integration of solar photovoltaic (PV) and biogas energy sources, complemented by a state-of-the-art energy storage system.\par
This tri-level optimization model is a testament to the commitment to precision and efficacy in addressing the complexities inherent in the operation of hybrid systems. By considering three distinct levels of objectives and their associated constraints, the model not only optimizes individual components but also orchestrates their interaction to achieve an unprecedented level of synergy. This holistic approach ensures that the grid-connected HRES operates at peak efficiency, maximizing energy production while minimizing environmental impact.\par
The incorporation of solar PV and biogas sources capitalizes on the strengths of each, providing a diversified and reliable energy supply. Solar PV, harnessing the power of sunlight, offers clean and sustainable energy, while biogas, derived from organic waste, contributes to waste management and provides a consistent energy source. The integration of these renewable sources is further bolstered by an advanced energy storage system, enhancing the system's overall reliability and responsiveness to fluctuating energy demands.\par
As a result, this synergistic combination has the potential to redefine the energy landscape, setting new benchmarks for sustainability, resilience, and efficiency. Beyond the technical intricacies, our tri-level optimization model signifies a forward-looking approach, contributing to the broader discourse on the role of advanced optimization techniques in shaping the future of energy systems. This endeavor aligns with the broader global commitment to fostering innovative solutions that pave the way towards a cleaner and more sustainable energy future \par
Within the confines of our meticulous modeling framework, a myriad of critical factors takes center stage, each carefully considered to craft a holistic approach. Our focus extends across crucial facets, encompassing the nuanced performance metrics of the solar photovoltaic (PV) array, the effectiveness of the energy storage system, and the imperative task of mitigating greenhouse gas emissions. In the pursuit of our objectives, we aspire to develop models that transcend the mere optimization of power feasibility; our ambition extends to fortifying system reliability while concurrently minimizing the environmental footprint.\par
Our guiding principle is to strike a harmonious balance—a symbiosis—between energy efficiency, operational dependability, and environmental sustainability. By intricately weaving these elements into our modeling fabric, we aim to create a robust foundation for the design and operation of modern energy systems.\par
These models, born out of a conscientious consideration of multiple dimensions, present a comprehensive approach to tackle the intricate challenges that characterize contemporary energy landscapes. Through the optimization of the seamless integration of diverse renewable sources, the augmentation of power efficiency, and an unwavering commitment to environmental sustainability, we actively contribute to the collective endeavor of shaping a greener and more resilient energy future.\par
As we navigate the complexities of our energy transition, the importance of our modeling framework becomes evident. It not only addresses the immediate needs of optimizing energy systems but also lays the groundwork for a future where energy efficiency, reliability, and environmental consciousness converge seamlessly. The harmonious balance we seek to achieve is not just a theoretical construct; it is a pragmatic pursuit that aligns with the aspirations of a world striving for sustainable energy solutions.\par
In summary, our modeling endeavors transcend the mere technical realm; they encapsulate a commitment to a holistic, balanced, and sustainable energy paradigm. By actively engaging with the intricate dance of optimizing renewable sources, we contribute meaningfully to the ongoing narrative of a cleaner, more efficient, and environmentally responsible energy landscape.\par
A tri-Level optimization model, as employed in this study, is a mathematical framework encompassing three objective functions operating at distinct hierarchical levels. The primary level, denoted as the leader, assumes responsibility for initial decision-making, while subsequent levels, known as followers, make decisions guided by their higher-level counterparts. This hierarchical model provides a structured method for addressing complex problems by breaking them down into manageable tiers of decision-making \cite{104}. \par

The proposed tri-level optimization model comprises four distinct levels, each with its own unique objective:\\
\textbf{Solar PV Efficiency (SPVE):} \\
This level seeks to maximize the efficiency of solar photovoltaic systems.\\
\textbf{Energy Storage System Performance (ESSP):} \\
Here, the aim is to maximize the performance of the energy storage system.\\
\textbf{Greenhouse Gas Emissions (GGE):}\\
 The objective is to minimize greenhouse gas emissions.\\
This multifaceted tri-level approach ensures a comprehensive and thorough examination of the entire system. By incorporating distinct levels dedicated to optimizing solar PV efficiency, energy storage system performance, and minimizing greenhouse gas emissions, we embrace a holistic perspective that spans efficiency, economic considerations, and environmental impact.\par
The presented mathematical model, designed as a multi-level formulation, is elegantly represented through equations 1 to 7. These equations encapsulate the intricate relationships and dependencies among the variables associated with each level, reflecting the interconnected nature of the system. As we delve into the mathematical representation, we illuminate the complex interplay between objectives, constraints, and the dynamic components of the hybrid renewable energy system.

\begin{equation}
%\begin{aligned}
\max \sum SPVE
\label{E1}
%\end{aligned}
\end{equation}
\begin{equation}
%\begin{aligned}
s.t. \;\;\;\;\;\;Constraints \;\;of \;\;Solar \;PV \;Efficiency\
\label{E2}
%\end{aligned}
\end{equation}
\begin{equation}
%\begin{aligned}
\max \sum ESSP
\label{E3}
%\end{aligned}
\end{equation}
\begin{equation}
%\begin{aligned}
s.t. \;\;\;\;\;\;Constraints \;\;of \;\; Energy \;Storage \;System\
\label{E4}
%\end{aligned}
\end{equation}
\begin{equation}
%\begin{aligned}
\min \sum GGE 
\label{E7}
%\end{aligned}
\end{equation}
\begin{equation}
%\begin{aligned}
s.t. \;\;\;\;\;\;Constraints \;\;of \;\; Greenhouse \;Gas \;Emissions
\label{E8}
%\end{aligned}
\end{equation}
\begin{equation}
%\begin{aligned}
s.t. \;\;\;\;\;\;General \;Constraints \;\;of \;\; HRES
\label{E8}
%\end{aligned}
\end{equation}

\section{Conclusion}
In conclusion, this paper introduces an innovative optimization model tailored for HRESs. The multi-level model is designed to accommodate scenarios where objectives exhibit diverse levels of significance. The framework provides a robust foundation for extending and enhancing the understanding of complex decision-making processes in the context of renewable energy integration. The flexibility of the proposed model allows for nuanced considerations of objective importance, paving the way for further refinement and application in diverse real-world contexts.

\end{document}